\documentclass[runningheads]{llncs}
\usepackage{graphicx}
\usepackage{comment}
\usepackage{amsmath,amssymb} % define this before the line numbering.
\usepackage{color}

\usepackage{bm}
\usepackage{booktabs}
\usepackage{tablefootnote}
\usepackage{mathtools}
\usepackage{authblk}
\usepackage{booktabs}
\usepackage{mwe}
\usepackage{graphicx}
\usepackage{subcaption}
\usepackage{textgreek}
\usepackage{moresize}
\usepackage{algorithm}
\usepackage{algpseudocode}
\usepackage{subcaption}
\usepackage[title]{appendix}
\usepackage[pagebackref=true,breaklinks=true,colorlinks,bookmarks=false]{hyperref}

\DeclarePairedDelimiterX{\infdivx}[2]{[}{]}{%
  #1\;\delimsize\|\;#2%
}
\newcommand{\kl}{\text{KL}\infdivx}

\makeatletter
\newcommand{\@chapapp}{\relax}%
\makeatother

\begin{document}
% \renewcommand\thelinenumber{\color[rgb]{0.2,0.5,0.8}\normalfont\sffamily\scriptsize\arabic{linenumber}\color[rgb]{0,0,0}}
% \renewcommand\makeLineNumber{\hss\thelinenumber\ \hspace{6mm} \rlap{\hskip\textwidth\ \hspace{6.5mm}\thelinenumber}}
% \linenumbers
\pagestyle{headings}
\mainmatter

\title{We Know Where We Don't Know: 3D Bayesian CNNs for Credible Geometric Uncertainty}

% CAMERA READY SUBMISSION
\titlerunning{3D Bayesian CNNs for Credible Geometric Uncertainty}
% If the paper title is too long for the running head, you can set
% an abbreviated paper title here
%
\author{Tyler LaBonte\inst{1,2} \and
Carianne Martinez\inst{1} \and
Scott A. Roberts\inst{1}}
\authorrunning{T. LaBonte \textit{et al.}}
% First names are abbreviated in the running head.
% If there are more than two authors, 'et al.' is used.
%
\institute{Sandia National Laboratories, Albuquerque, New Mexico, USA\thanks{\ssmall{This paper describes objective technical results and analysis. Any subjective views or opinions that might be expressed in the paper do not necessarily represent the views of the U.S. Department of Energy or the United States Government. Supported by the Laboratory Directed Research and Development program at Sandia National Laboratories, a multimission laboratory managed and operated by National Technology and Engineering Solutions of Sandia, LLC., a wholly owned subsidiary of Honeywell International, Inc., for the U.S. Department of Energy’s National Nuclear Security Administration under contract DE-NA-0003525. This manuscript has been authored by National Technology \& Engineering Solutions of Sandia, LLC. under Contract No. DE-NA0003525 with the U.S. Department of Energy/National Nuclear Security Administration. The United States Government retains and the publisher, by accepting the article for publication, acknowledges that the United States Government retains a non-exclusive, paid-up, irrevocable, world-wide license to publish or reproduce the published form of this manuscript, or allow others to do so, for United States Government purposes. SAND2020-3269 R.}} \and
University of Southern California, Los Angeles, California, USA
\email{tlabonte@usc.edu, \{cmarti5, sarober\}@sandia.gov}}

\maketitle

\begin{abstract}
Deep learning has been successfully applied to the segmentation of 3D Computed Tomography (CT) scans. Establishing the credibility of these segmentations requires uncertainty quantification (UQ) to identify untrustworthy predictions. Recent UQ architectures include Monte Carlo dropout networks (MCDNs), which approximate deep Gaussian processes, and Bayesian neural networks (BNNs), which learn the distribution of the weight space. BNNs are advantageous over MCDNs for UQ but are thought to be computationally infeasible in high dimension, and neither architecture has produced interpretable geometric uncertainty maps. We propose a novel 3D Bayesian convolutional neural network (BCNN), the first deep learning method which generates statistically credible geometric uncertainty maps and scales for application to 3D data. We present experimental results on CT scans of graphite electrodes and laser-welded metals and show that our BCNN outperforms an MCDN in recent uncertainty metrics. The geometric uncertainty maps generated by our BCNN capture distributions of sigmoid values that are interpretable as confidence intervals, critical for applications that rely on deep learning for high-consequence decisions. Code available at \url{https://github.com/sandialabs/bcnn}.
\keywords{Uncertainty quantification, volumetric segmentation, variational inference}
\end{abstract}

\begin{figure}[t!]
        \centering
         \begin{subfigure}[t]{0.35\textwidth}
            \centering
            \includegraphics[width=0.95\textwidth]{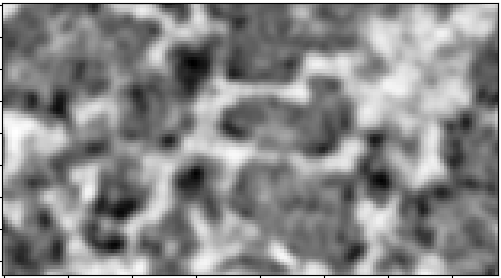}
            \caption[]%
            {{\small CT scan slice.}}    
        \end{subfigure}
        \hskip0.2em
        \begin{subfigure}[t]{0.35\textwidth}  
            \centering 
            \includegraphics[width=0.95\textwidth]{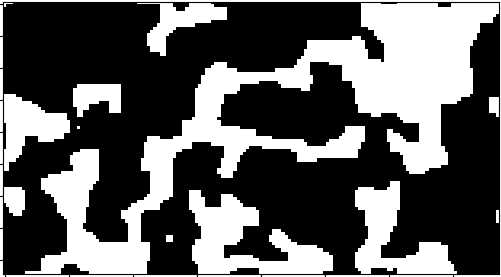}
            \caption[]%
            {{\small Target segmentation.}}    
        \end{subfigure}
        \vskip0.33\baselineskip
	\hskip0.6em
        \begin{subfigure}[t]{0.35\textwidth}
            \centering
            \includegraphics[width=\textwidth]{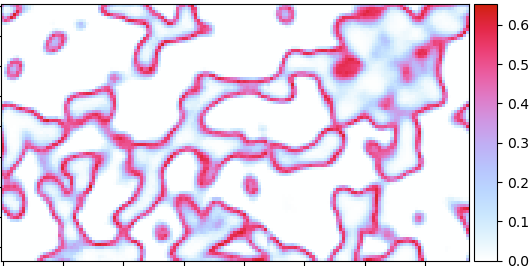}
            \caption[]%
            {{\small BCNN uncertainty.}}    
        \end{subfigure}
        \hskip0.25em
        \begin{subfigure}[t]{0.35\textwidth}  
            \centering 
            \includegraphics[width=\textwidth]{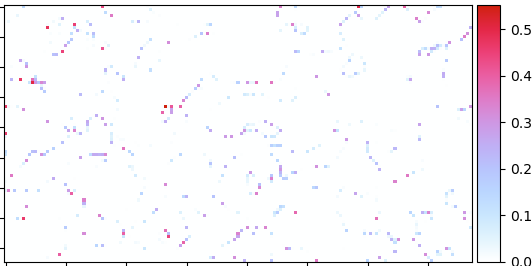}
            \caption[]%
            {{\small MCDN uncertainty.}}    
        \end{subfigure}
        \caption{Zoomed Uncertainty Maps on Graphite Test Set Sample III, Slice 64. Note that our BCNN uncertainty map captures continuity and visual gradients while the MCDN uncertainty map is pixelated and uninterpretable.}
        \label{fig:iii_zoom}
\end{figure}

\section{Introduction}

Non-destructive 3D imaging techniques allow scientists to study the interior of objects which cannot otherwise be observed. In particular, 3D Computed Tomography (CT) scans are used in manufacturing to identify defects before a part is deployed and to certify physical properties of materials. A critical step in the analysis of CT scans is segmentation, wherein an analyst labels each voxel in a scan (\textit{e.g.,} as material or void). However, due to the noise and artifacts found in CT scans along with human error, these segmentations are expensive, irreproducible, and unreliable \cite{Martinez}. Deep learning models such as convolutional neural networks (CNNs) have revolutionized the automated segmentation of 3D imaging by providing a fast, accurate solution to these challenges \cite{Milletari,Cicek}.

For use with high-consequence part certification, \textit{e.g.,} in automobiles and aircraft, segmentation must include uncertainty quantification (UQ) to obtain accurate safety confidence intervals. Recent research casts deep neural networks as probabilistic models in order to obtain uncertainty measurements. Two common UQ architectures are Monte Carlo dropout networks (MCDNs) \cite{Gal} and variational inference-based Bayesian neural networks (BNNs) \cite{Blundell}. MCDNs enable UQ in the output space with little computational cost \cite{Gal}, but their statistical validity has been questioned \cite{Osband}. In contrast, BNNs measure uncertainty in the weight space, resulting in statistically-justified UQ at the cost of at least double the number of trainable parameters \cite{Gal}, increased convergence time, and sensitivity to hyperparameter optimization \cite{Ovadia}.

To the best of our knowledge there is no existing deep learning model that successfully generates statistically credible geometric uncertainty maps, in either 2D or 3D. Recent work has theorized that this is because BNNs are not scalable \cite{Gal,Lak}. We refute this and propose a novel 3D Bayesian CNN (BCNN) architecture, the first variational inference-based architecture which enables segmentation and UQ in a 3D domain. Our BCNN effectively predicts binary segmentations of billion-voxel CT scans and generates credible geometric uncertainty maps which the MCDN cannot capture. We show via experimental results on CT scan datasets of graphite electrodes and laser-welded metals that our BCNN outperforms the MCDN on recent uncertainty metrics \cite{Mukhoti}.

The major advantage of our BCNN is that it measures uncertainty in the \textit{weight space}, while the MCDN measures uncertainty in the \textit{output space}. It is possible for MCDNs to provide uncertainty maps, but to obtain credible UQ, we must study the distribution of the weight space. Because our BCNN measures this distribution, it provides meaningful uncertainty maps that can be directly interpreted as confidence intervals on the segmentation. As shown in Figure \ref{fig:iii_zoom}, our BCNN uncertainty maps capture continuity and visual gradients, which is a major advantage not only for material simulations as discussed in Section \ref{material-advantages}, but for any application where geometric uncertainty must be quantified and understood. While we have leveraged many ideas from previous work, our contribution is not a straightforward extension from 2D to 3D, but instead a discovery of a unique synthesis of techniques that enabled successful training and segmentation of large 3D volumes with credible geometric UQ.

In summary, our contributions are:
\begin{enumerate}
\item The first deep learning method which generates statistically credible and interpretable geometric uncertainty maps.
\item The first Bayesian CNN that scales for use with three-dimensional data.
\item A computationally feasible pipeline for generating and interpreting geometric uncertainty for use as confidence intervals around segmentation.
\item Open-source code for volumetric segmentation with UQ.
\end{enumerate}

\section{Related Work}

In this section, we describe recent publications in volumetric segmentation and UQ which enabled the success of our BCNN.

\subsection{Volumetric Segmentation}

The problem of volumetric segmentation has seen much high-impact work in the past few years. The 2D fully convolutional network \cite{Long} and U-Net \cite{Ronneberger} led \cite{Milletari} to propose the first 3D CNN for binary segmentation of MRI images, called V-Net. Closely afterward, \cite{Cicek} proposed 3D U-Net, a direct extension of the U-Net to a 3D domain. While V-Net was designed for binary segmentation of the human prostate and 3D U-Net was designed for binary segmentation of the kidney of the \textit{Xenopus}, they both employ an encoder-decoder architecture inspired by U-Net \cite{Milletari,Cicek}. In this technique, a 3D volume is mapped to a latent space via successive convolutional and pooling layers; this latent representation is then upsampled and convolved until it reaches the size of the original volume and and generates a per-voxel segmentation \cite{Ronneberger}.

While most volumetric segmentation work pertains to medicine, 3D materials segmentation is also an active area of research due to the importance of quality segmentations in physics simulations. In particular, \cite{Kono} employed fully convolutional networks to segment CT scan volumes of short glass fibers achieving the first accurate results in low-resolution fiber segmentation, and \cite{MacNeil2019} proposed a semi-supervised algorithm for segmentation of woven carbon fiber volumes from sparse input.

\subsection{Uncertainty Quantification}

While deep learning models often outperform traditional statistical approaches in terms of accuracy and generalizability, they do not have built-in uncertainty measurements like their statistical counterparts. \cite{Gal} showed that predictive probabilities (\textit{i.e.,} the softmax outputs of a model) are often erroneously interpreted as an uncertainty metric. Instead, recent work has cast neural networks as Bayesian models via approximating probabilistic models \cite{Gal} or utilized variational inference to learn the posterior distribution of the neural network weights \cite{Blundell}.

\subsubsection{Monte Carlo Dropout Networks (MCDNs)}

\cite{Gal} showed that a neural network with dropout applied before every weight layer (\textit{i.e.,} an MCDN) is mathematically equivalent to an approximation to a deep Gaussian process \cite{Damianou}. Specifically, one can approximate a deep Gaussian process with covariance function \( \mathbf{K}(\bm{x}, \bm{y}) \) by placing a variational distribution over each component of a spectral decomposition of \( \mathbf{K} \). This maps each layer of the deep Gaussian process to a layer of hidden units in a neural network. By averaging stochastic forward passes through the dropout network at inference time, one obtains a Monte Carlo approximation of the intractable approximate predictive distribution of the deep Gaussian process \cite{Gal}; thus the voxel-wise standard deviations of the predictions are usable as an uncertainty metric.

One of the top benefits of the MCDN is its ease of implementation; as an architecture-agnostic technique which is dependent only on the dropout layers, Monte Carlo dropout can easily be added to very large networks without an increase in parameters. As a result, MCDNs have been implemented with good results in several different applications. In particular, \cite{Liu} successfully implemented a 3D MCDN for UQ in binary segmentations of MRI scans of the amygdala, and \cite{Martinez} used V-Net \cite{Milletari} with Monte Carlo dropout for UQ in binary segmentations of CT scans of woven composite materials.

While the MCDN is one of the most common UQ architectures used in deep learning, its statistical soundness has been called into question. \cite{Osband} argues that Monte Carlo dropout provides an approximation to the risk of a model rather than its uncertainty (in other words, that it approximates the inherent stochasticity of the model rather than the variability of the model's posterior belief). \cite{Osband} also shows that the posterior distribution given by dropout does not necessarily converge as more data is gathered; instead, the posterior depends only on the interaction between the dropout rate and the model size.

\subsubsection{Bayesian Neural Networks (BNNs)}

Another approach to UQ in deep neural networks is Bayesian learning via variational inference (\textit{i.e.,} a BNN). Instead of point estimates, the network learns the posterior distribution over the weights given the dataset, denoted \( P(\bm{w}|\mathcal{D}) \), given the prior distribution \( P(\bm{w}) \). However, calculating the exact posterior distribution is intractable due to the extreme overparametrization found in neural networks \cite{Blundell}. Previous work by \cite{Hinton} and \cite{Graves} proposed variational learning as a method to approximate the posterior distribution. Variational learning finds the parameters \( \theta \) of the distribution \( q(\bm{w}|\theta) \) via the minimization of the variational free energy cost function, often called the expected lower bound (ELBO). It consists of the sum of the Kullback-Leibler (KL) divergence and the negative log-likelihood (NLL), which \cite{Blundell} explains as a tradeoff between satisfying the simplicity prior (represented by the KL term) and satisfying the complexity of the dataset (represented by the NLL term):
\begin{equation}
\label{eq:vfe}
\mathcal{F}(\mathcal{D},\theta) = \kl{q(\bm{w}|\theta)}{P(\bm{w})} - \mathbb{E}_{q(\bm{w}|\theta)}[\log P(\mathcal{D}|\bm{w})].
\end{equation}

\cite{Blundell} proposed the Bayes by Backprop algorithm, which combines variational inference with traditional backpropagation to efficiently find the best approximation to the posterior. Bayes by Backprop uses the gradients calculated in backpropagation to ``scale and shift" the variational parameters of the posterior, updating the posterior with minimal additional computation \cite{Blundell}. 

One challenge associated with probabilistic weights is that all examples in a mini-batch typically have similarly sampled weights, limiting the variance reduction effect of large mini-batches \cite{Wen}. \cite{Kingma} introduced local reparametrization, which greatly reduces the variance of stochastically sampled weights by transforming global weight uncertainty into independent local noise across examples in the mini-batch. In a similar vein, \cite{Wen} proposed the Flipout estimator, which empirically achieves ideal variance reduction by sampling weights pseudo-independently for each example. While local reparametrization only works for fully-connected networks, Flipout can be used effectively in fully-connected, convolutional, and recurrent networks \cite{Wen}.

In the 2D domain, \cite{Shridhar} proposed a 2D BCNN that extended local reparametrization to CNNs but did not generate geometric uncertainty maps. \cite{Ovadia} showed that 2D BCNNs with Flipout \cite{Wen} are effective for non-geometric UQ on the MNIST and CIFAR-10 datasets, but they found it was difficult to get BCNNs to work with complex datasets, supporting the hypothesis that BNNs are not scalable \cite{Gal,Lak} which we refute with our 3D BCNN.

\begin{figure*}
\begin{center}
\includegraphics[scale=0.2]{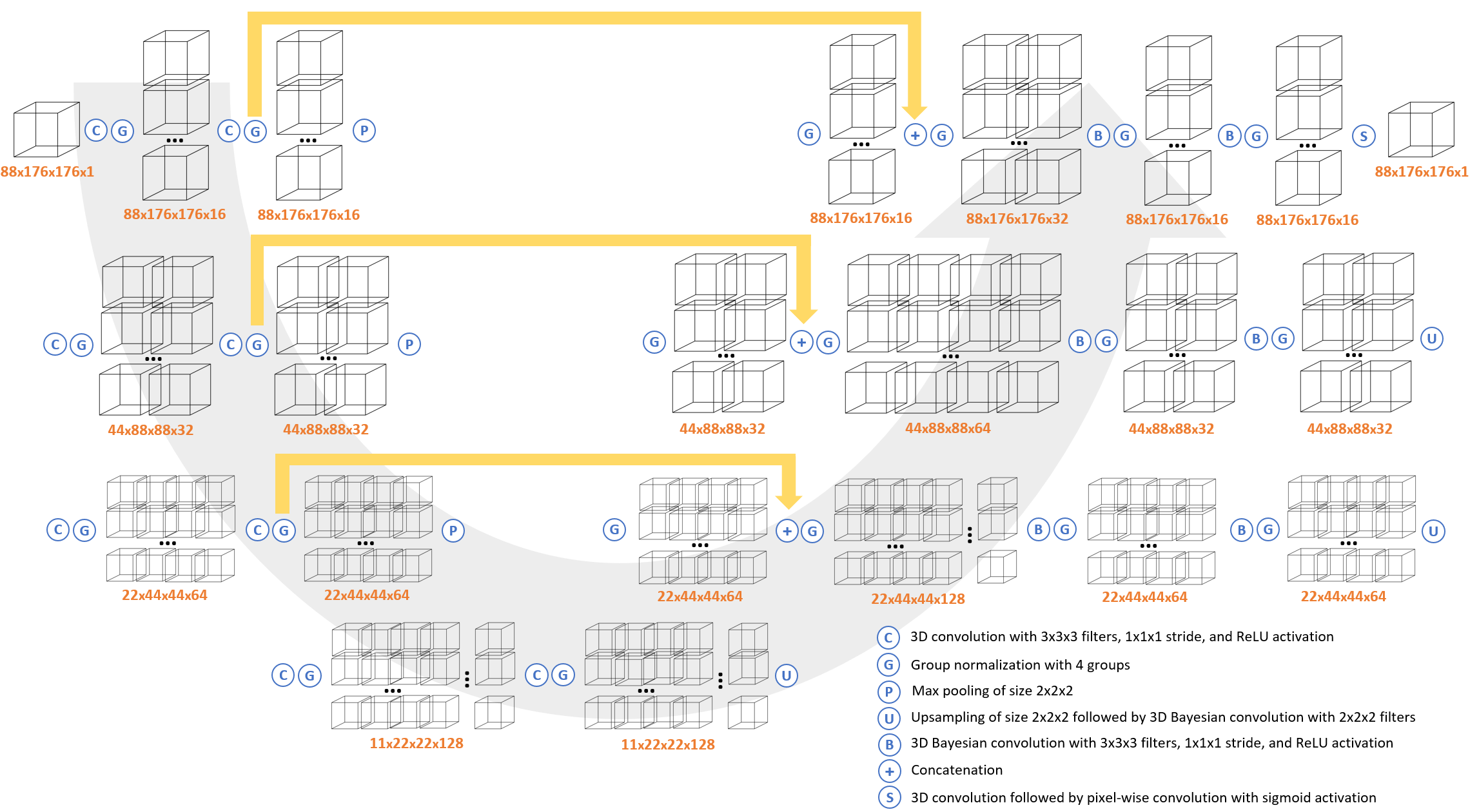}
\end{center}
   \caption{Schematic of our BCNN architecture with sample volume dimensions from the Graphite dataset. Measurements are (depth, height, width, channels). Best viewed in electronic format.}
   \label{fig:bcnn}
\end{figure*}

\section{Methodology}

In this section, we present our BCNN architecture and describe our reasoning behind several design decisions.

\subsection{Architecture}

In Figure \ref{fig:bcnn}, we present a schematic representation of our BCNN architecture. Similarly to V-Net \cite{Milletari}, we employ an encoder-decoder architecture. The encoder half (left) of the network compresses the input into a latent space while the decoder half (right) decompresses the latent representation of the input into a segmentation map. We do not include stochastic layers in the encoder half of the network to maximize the amount of information transfer between the original volume and the latent space.

The encoder half of the network has four stages, each with two convolutional layers and normalization layers followed by a max pooling layer to reduce the size of the input. Thus, after each layer, the volume's depth, height, and width are halved while its channels are doubled, reducing the size of the volume by a factor of four.

The decoder half of the network consists of three stages, corresponding to the first three layers of the encoder half. First, we upsample the output of the previous layer and apply convolutional and normalization layers to double the volume's depth, height, and width while halving its channels. We then concatenate this volume with the pre-pooling output of the corresponding encoder layer; this skip connection assists in feature-forwarding through the network. Then, we apply two more convolutional and normalization layers, a final convolutional layer, and a sigmoid activation. This results in a volume of the same size as the input representing a binary segmentation probability map.

In the decoder half of the network, we implement volumetric convolutional layers with distributions over the weights. Each Bayesian convolutional layer is initialized with a standard normal prior \(P(\bm{w}) =\mathcal{N}(0,1)\) and employs the aforementioned Flipout estimator \cite{Wen} to approximate the distribution during forward passes. Our implementation draws from the Bayesian Layers library \cite{Tran} included in TensorFlow Probability \cite{Dillon}, which monitors the KL divergence of the layer's posterior distribution with respect to its prior. Our BCNN has 1,924,964 trainable parameters, while its MCDN counterpart has 1,403,059.

\subsection{Design Decisions}

Since training volumes can be quite large, our batch size is constrained by the amount of available GPU memory, resulting in a batch size too small for batch normalization to accurately compute batch statistics. Thus, we utilize group normalization \cite{Wu}, which normalizes groups of channels and is shown to have accurate performance independent of batch size. Proper normalization was observed to be a critical factor in the convergence of our model; by tuning the number of groups used in the group normalization layers, we found that our model converged most reliably when using four groups.

At each downward layer \( i \), we apply \( 2^{3+i} \) filters. This was found to be more effective than a more simple model with \( 2^{2+i} \) filters and a more complex model with \( 2^{4+i} \) filters. We hypothesize that some minimum amount of learned parameters was necessary to produce accurate segmentations, but with \( 2^{4+i} \) filters, the model's overparameterization made training significantly more difficult.

We tested many prior distributions, including scale mixture \cite{Blundell}, spike-and-slab \cite{Mitchell}, and a normal distribution with increased variance, but found that a standard normal prior provided the best balance between weight initialization and weight exploration. Skip connections were found to slightly increase the accuracy of our predictions by forwarding fine-grained features that otherwise would have been lost in the encoder half of the network. We tested both max pooling and downward convolutional layers and observed negligible difference.

\section{Experiments}

In this section, we describe our datasets and detail our procedures.

\subsection{Datasets}

Two 3D imaging datasets are used to test our BCNN. The first is a series of CT scans of graphite electrodes for lithium-ion batteries, which we refer to as the Graphite dataset \cite{Mueller2018,Pietsch2018}. This material consists of non-spherical particles (dark objects in the images) that are coated onto a substrate and calendared to densify. The academically manufactured (``numbered'') electrodes \cite{Mueller2018} were imaged with 325 nm resolution and a domain size of \( 700 \times 700 \times (48-75) \) \textmu m. The commercial (``named'') electrodes \cite{Pietsch2018} were imaged at 162.5 nm resolution and a domain size of \( 416 \times 195 \times 195 \) \textmu m. Eight samples were studied, each with 500 million to 1 billion voxels. Each volume was hand-segmented using commercial tools \cite{Norris2020}; these manual segmentations were used for training and testing. We trained our BCNN on the GCA400 volume and tested on the remaining seven electrodes.

Laser-welded metal joints comprise a second dataset, which we refer to as the Laser Weld dataset. To generate these volumes, two metal pieces are put into contact and joined with an incident laser beam. The light regions of the resulting scans represent voids or defects in the weld. The Laser Weld dataset consists of CT scans of ten laser-welded metal joints, each with tens of millions of voxels. Similarly to the Graphite dataset, these volumes were hand-segmented \cite{Norris2020}. We trained a separate BCNN on samples S2, S24, and S25, then tested on the remaining seven held-out volumes.

For both datasets, we normalized each CT scan to have voxel values with zero mean and unit variance. Additionally, each CT scan was large enough to require that we process subvolumes of the 3D image rather than ingesting the entire scan as a whole into the neural network on the GPU. Our algorithm for preprocessing these volumes is set forth in the Appendix.

\subsection{Training}

Recall the variational free energy loss \cite{Blundell} from Equation \ref{eq:vfe}. We utilize the amendment \cite{Graves} to mini-batch optimization for mini-batch \( i\in \{1,2,\dots,M\} \) by dividing the KL term by \(M\). This distributes the KL divergence penalty evenly over each minibatch; without this scaling, the KL divergence term dominates the equation, causing the model to converge to a posterior with suboptimal accuracy.

We also use monotonic KL annealing \cite{Bowman} as detailed in Equation \ref{eq:an}; this annealing was necessary for the reliable convergence of our model as it allowed the model to learn the 3D segmentation before applying the KL divergence penalty. We denote the current epoch as \(E\) and accept as hyperparameters a KL starting epoch \(s\), initial KL weight \(k_0\), and step value \(k_1\) to obtain the KL weight for the current epoch \(k_E\) as follows:
\begin{equation}
\label{eq:an}
k_E =
\begin{cases}
k_0 & \text{if } E \leq s \\
\min(1, k_0 + k_1(E - s)) & \text{if } E > s
\end{cases}
\end{equation}
For the Graphite dataset we use \(s=1, k_0=1/2,k_1=1/2\) and for the Laser Weld dataset we use \(s=1, k_0=0,k_1=1/4\). We use the aforementioned Bayes by Backprop algorithm \cite{Blundell} to train our BCNN under the resultant loss function:
\begin{equation}
\mathcal{F}^{E}_i(\mathcal{D}_i,\theta) = \frac{k_E}{M}\kl{q(\bm{w}|\theta)}{P(\bm{w})} - \mathbb{E}_{q(\bm{w}|\theta)}[\log P(\mathcal{D}_i|\bm{w})].
\end{equation}
We use the Adam optimizer \cite{KingmaBa} with learning rate \( \alpha=0.0001 \) for the Graphite dataset and \( \alpha=0.001 \) for the Laser Weld dataset. We parallelized our model and trained on two NVIDIA Tesla V100 GPUs with 32GB of memory each. For our BCNN, one epoch of 1331 chunks of size \( 88 \times 176 \times 176 \) took approximately 17 minutes and 30 seconds with a maximum batch size of 3. We trained each model for 2 epochs on the 4913-sample Graphite dataset and for 7 epochs on the 549-sample Laser Weld dataset.

\subsection{Testing}

We computed 48 Monte Carlo samples on each test chunk to obtain a distribution of sigmoid values for each voxel. The Monte Carlo dropout technique is justified in representing uncertainty as the standard deviation of the sigmoid values because it approximates a deep Gaussian process \cite{Gal}; however, our BCNN does not guarantee adherence to a normal distribution in practice. Thus, in order to effectively compare the outputs of both networks while mimicking the standard deviation measurement of the MCDN, we represent confidence intervals on the segmentation as the 33\textsuperscript{rd} and the 67\textsuperscript{th} percentiles of the sigmoid values, and uncertainty as the difference. We compare our results against an MCDN of identical architecture except with regular convolutional layers instead of Bayesian convolutional layers and spatial dropout \cite{Tompson} applied at the end of each stage prior to upsampling.

\section{Results}

In this section, we present inference results of our BCNN and compare its performance with the MCDN.

\begin{figure}
        \centering
        \begin{subfigure}[t]{0.3\textwidth}
            \centering 
            \includegraphics[width=0.85\textwidth]{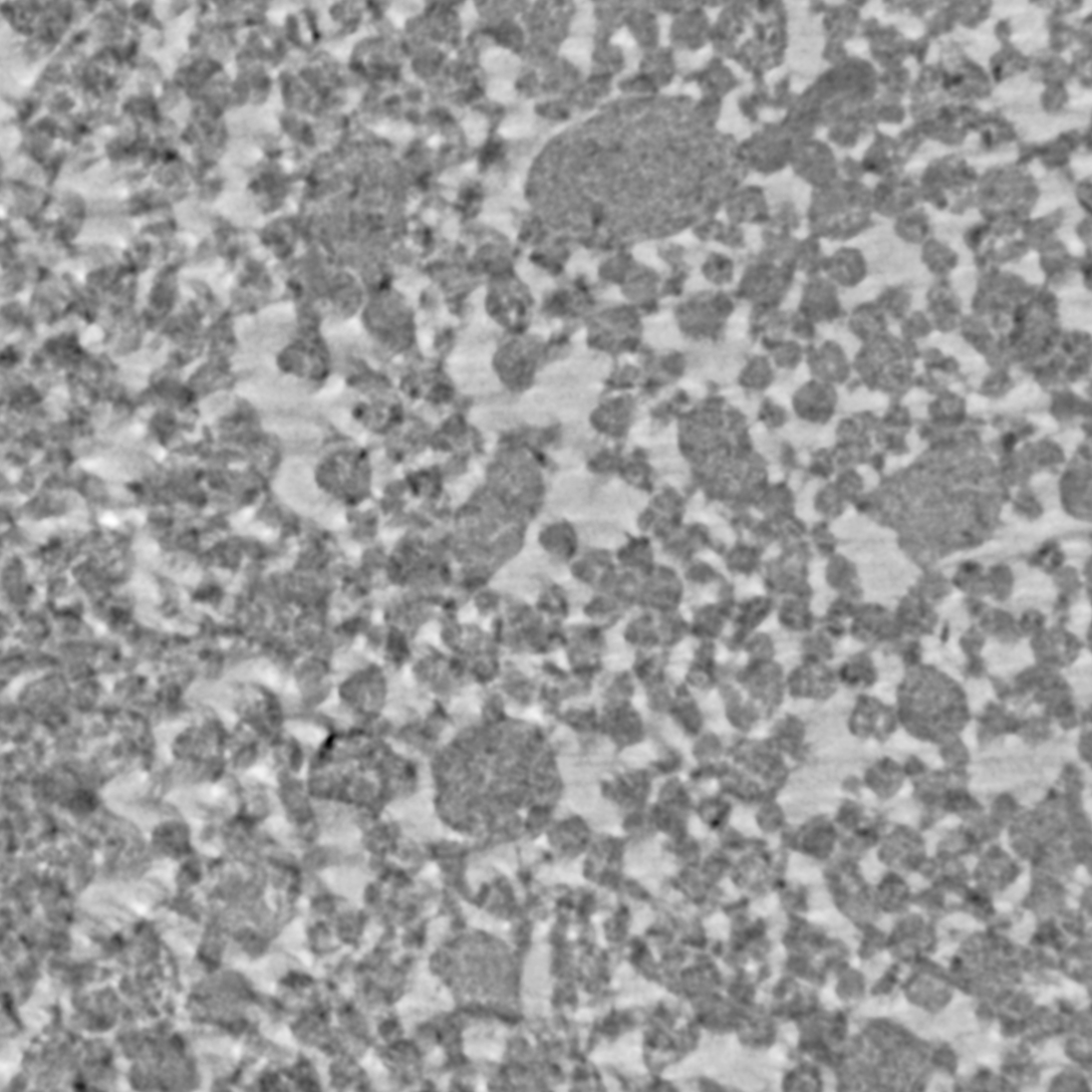}
            \caption[]%
            {{\small Test slice.}}
        \end{subfigure}
        \hskip0.2em
        \begin{subfigure}[t]{0.3\textwidth}  
	  \centering 
            \includegraphics[width=0.85\textwidth]{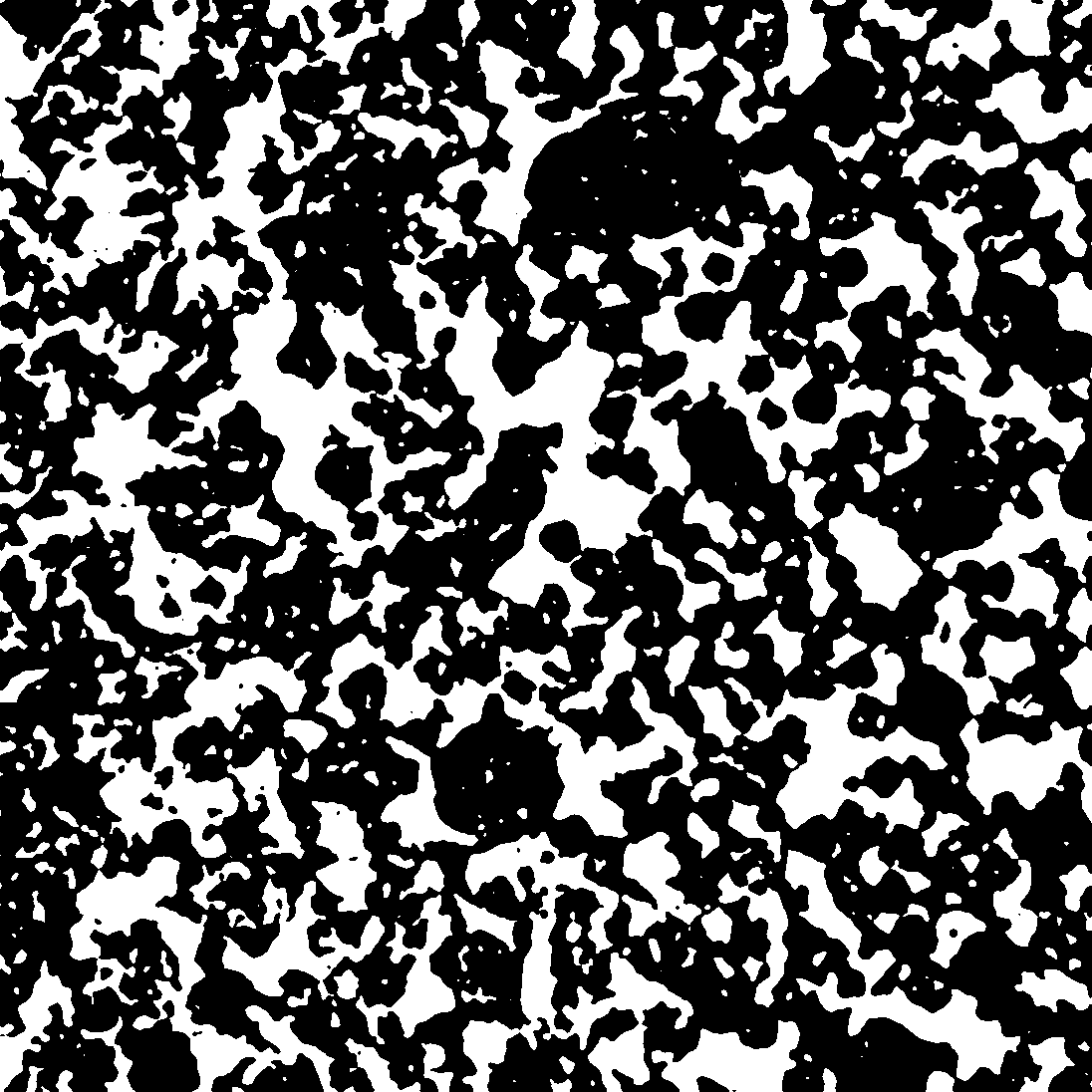}
            \caption[]%
            {{\small BCNN segmentation.}}
        \end{subfigure}
        \hskip0.2em
        \begin{subfigure}[t]{0.3\textwidth}    
\centering 
            \includegraphics[width=0.85\textwidth]{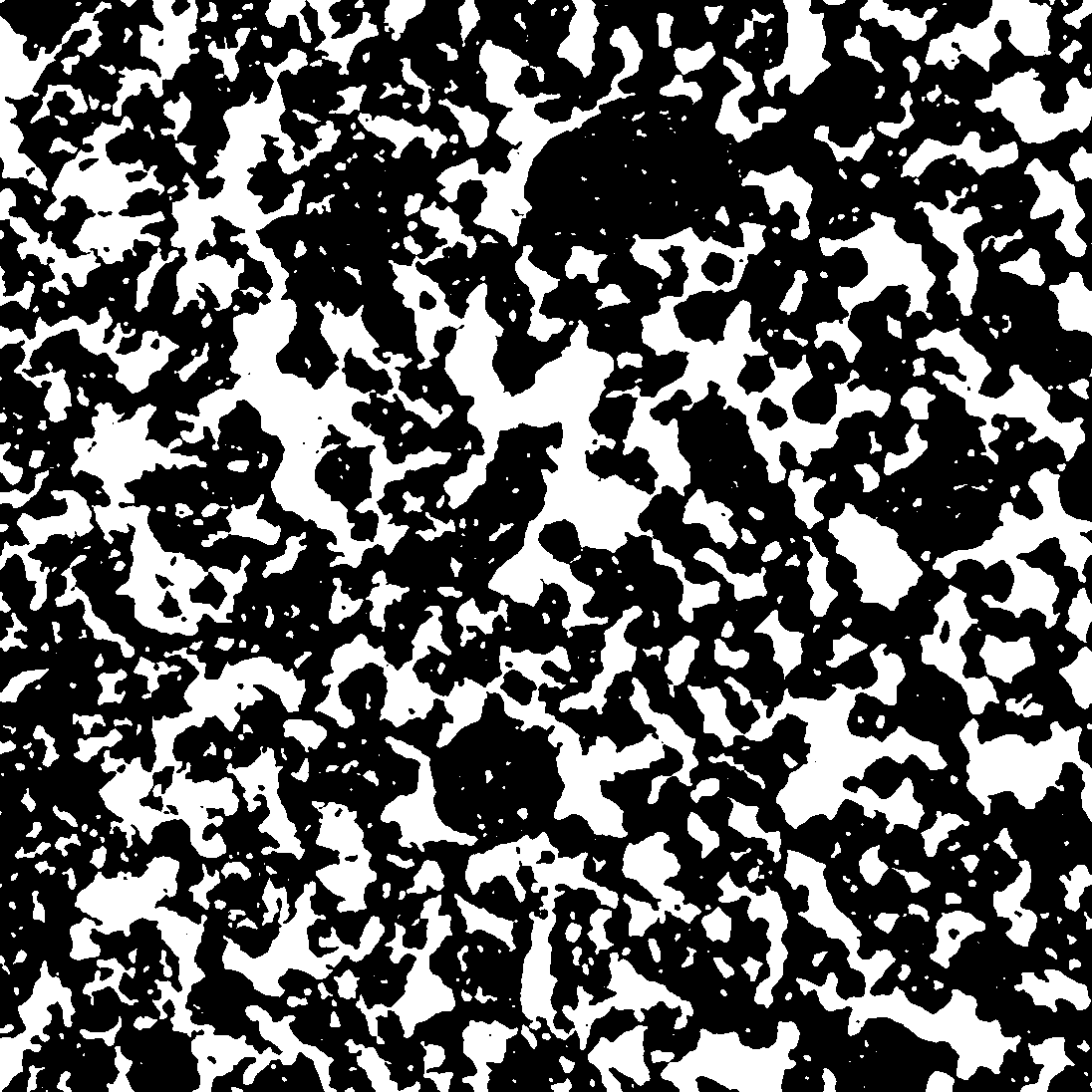}
            \caption[]%
            {{\small MCDN segmentation.}}   
        \end{subfigure}
        \vskip0.33\baselineskip
        \hskip0.5em
        \begin{subfigure}[t]{0.3\textwidth}
		\centering
            \includegraphics[width=0.85\textwidth]{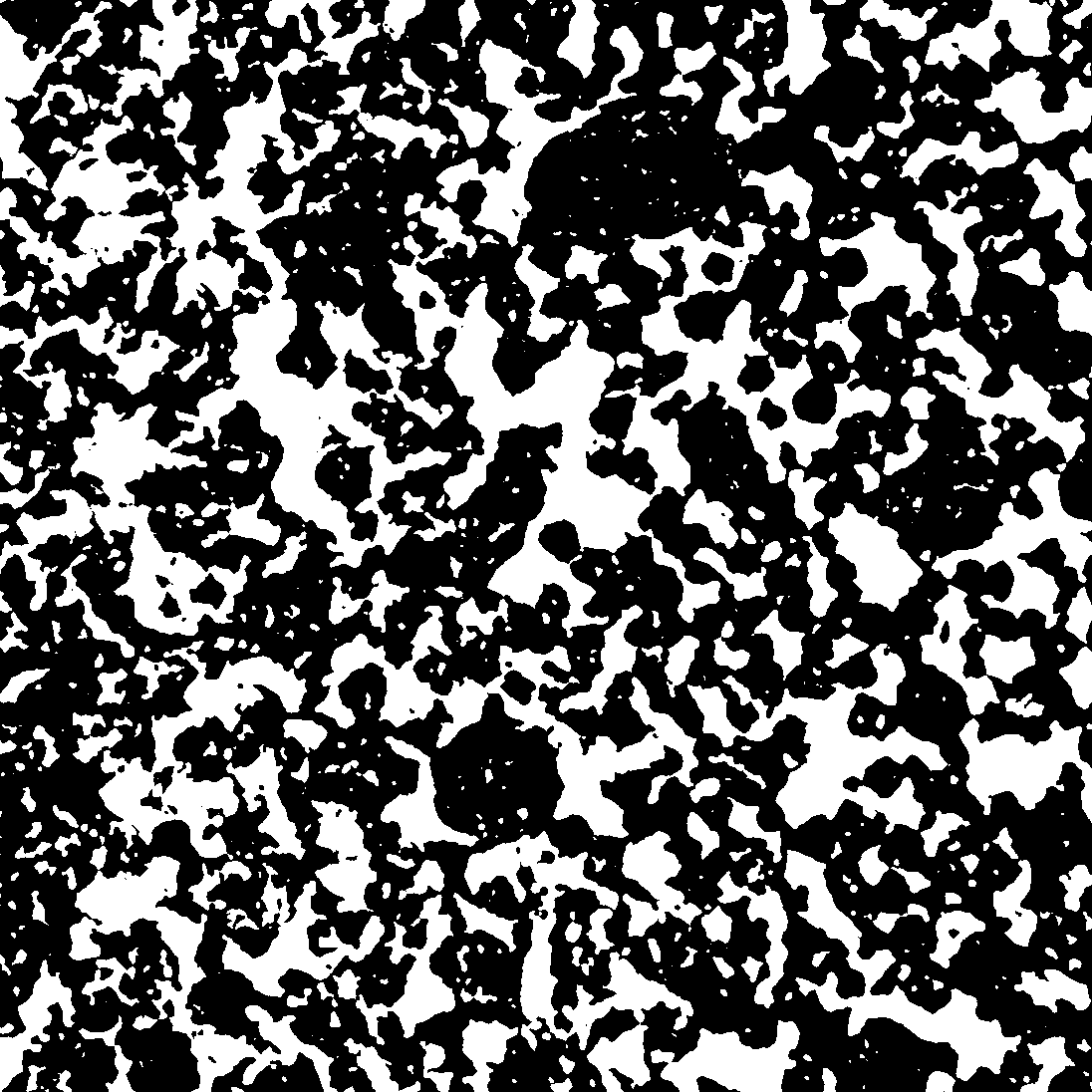}
            \caption[]%
            {{\small Target segmentation.}}
        \end{subfigure}
        \hskip0.9em
        \begin{subfigure}[t]{0.3\textwidth}   
            \centering 
            \includegraphics[width=0.98\textwidth]{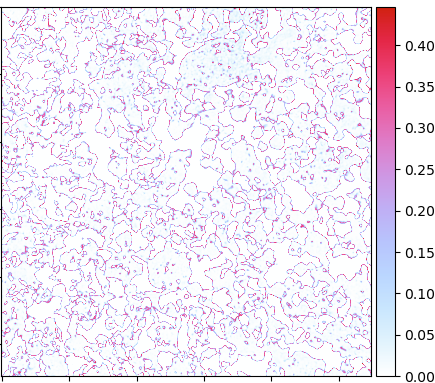}
            \caption[]%
            {{\small BCNN uncertainty.}}    
        \end{subfigure}
        \hskip0.2em
        \begin{subfigure}[t]{0.3\textwidth}  
            \centering 
            \includegraphics[width=0.98\textwidth]{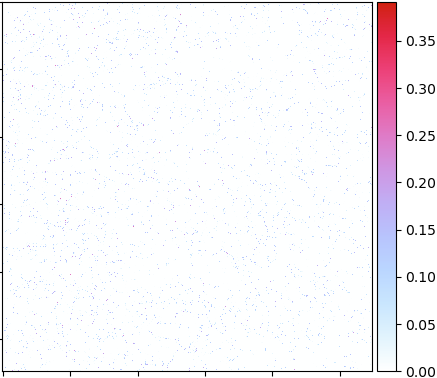}
            \caption[]%
            {{\small MCDN uncertainty.}}    
        \end{subfigure}
        \caption{Results on Graphite Test Set Sample GCA2K, Slice 212. Our BCNN produces interpretable uncertainty even in this complex, billion-voxel volume.}
        \label{fig:GCA2K}
\end{figure}

\subsection{Prediction Results}

Figure \ref{fig:GCA2K} shows a successful segmentation and uncertainty measurements on the billion-voxel GCA2K sample from the Graphite dataset. Our BCNN provides an equivalent or better segmentation than the MCDN and produces an interpretable geometric uncertainty map, showing that it can scale to large 3D domains. Figure \ref{fig:iii_zoom} shows a zoomed-in portion of the III sample uncertainty map which highlights the continuity and visual gradients captured in our BCNN uncertainty map, while the MCDN produces uninterpretable pixel-by-pixel uncertainty measurements. This is an advantage of our BCNN measuring the model uncertainty in the weight space, rather than in the output space like the MCDN.

\begin{figure}[h]
        \centering
        \begin{subfigure}[t]{0.3\textwidth}
	 \centering 
            \includegraphics[width=0.95\textwidth]{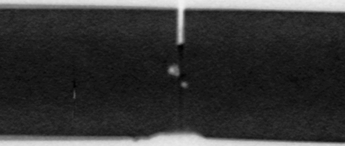}
            \caption[]%
            {{\small CT scan slice.}}      
        \end{subfigure}
        \hskip0.2em
        \begin{subfigure}[t]{0.3\textwidth}
\centering 
            \includegraphics[width=0.95\textwidth]{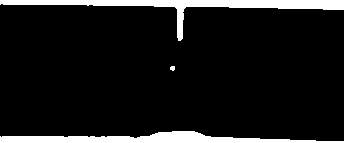}
            \caption[]%
            {{\small BCNN segmentation.}}
        \end{subfigure}
        \hskip0.2em
        \begin{subfigure}[t]{0.3\textwidth}
	    \centering
            \includegraphics[width=0.95\textwidth]{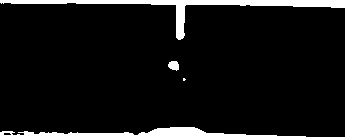}
            \caption[]%
            {{\small MCDN segmentation.}}		
        \end{subfigure}
        \vskip0.33\baselineskip
	\hskip0.1em
        \begin{subfigure}[t]{0.3\textwidth}
 	\centering
	    \raisebox{0.4\height}{
            \includegraphics[width=0.95\textwidth]{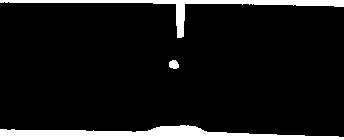}}
            \caption[]%
            {{\small Target segmentation.}} 
        \end{subfigure}  
        \hskip0.8em
        \begin{subfigure}[t]{0.3\textwidth}   
            \centering 
            \includegraphics[width=\textwidth]{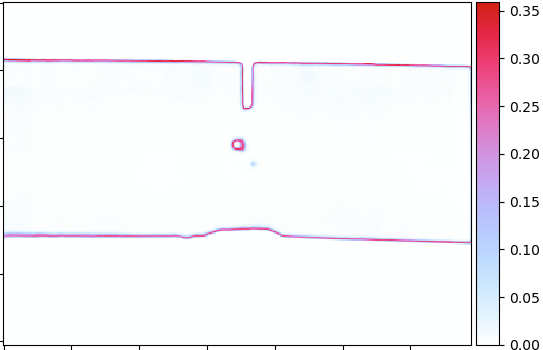}
            \caption[]%
            {{\small BCNN uncertainty.}}    
        \end{subfigure}
        \hskip0.3em
        \begin{subfigure}[t]{0.3\textwidth}  
            \centering 
            \includegraphics[width=\textwidth]{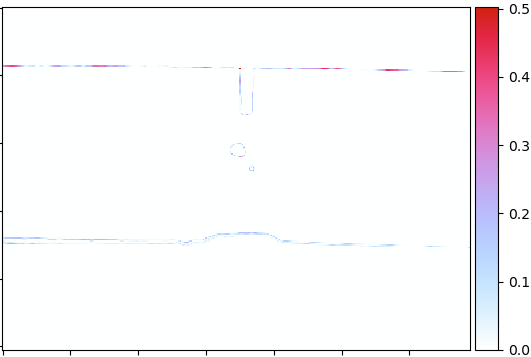}
            \caption[]%
            {{\small MCDN uncertainty.}}    
        \end{subfigure}
        \caption{Results on Laser Weld Test Set Sample S33, Slice 604. Our BCNN achieves a more accurate segmentation in addition to producing an uncertainty map with consistent uncertainty measurements across the borders of the weld. Additionally, our BCNN learned that there is a distribution of uncertainty around the central void, whereas the MCDN represents it with a pixel-wide line.}
        \label{fig:S33}
\end{figure}

\begin{figure}[h!]
        \centering
        \begin{subfigure}[t]{0.3\textwidth}
	 \centering 
            \includegraphics[width=0.9\textwidth]{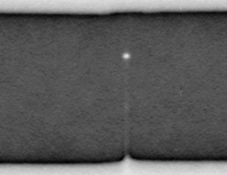}
            \caption[]%
            {{\small CT scan slice.}}    
        \end{subfigure}
        \hskip0.2em
        \begin{subfigure}[t]{0.3\textwidth}
	\centering
	\raisebox{0.05\height}{
            \includegraphics[width=0.9\textwidth]{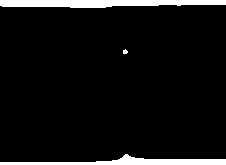}}
            \caption[]%
            {{\small BCNN segmentation.}}
        \end{subfigure}
        \hskip0.2em
        \begin{subfigure}[t]{0.3\textwidth}
	    \centering
	    \raisebox{0.05\height}{
            \includegraphics[width=0.9\textwidth]{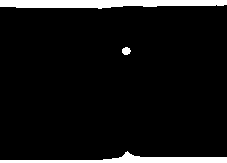}}
            \caption[]%
            {{\small MCDN segmentation.}}
        \end{subfigure}
        \vskip0.33\baselineskip
        \begin{subfigure}[t]{0.3\textwidth} 
 	    \centering
	    \raisebox{0.25\height}{
            \includegraphics[width=0.9\textwidth]{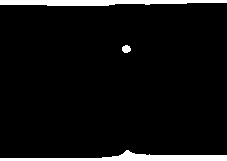}}
            \caption[]%
            {{\small Target segmentation.}}   
        \end{subfigure}
        \hskip0.8em
        \begin{subfigure}[t]{0.3\textwidth}
            \centering 
            \includegraphics[width=0.9\textwidth]{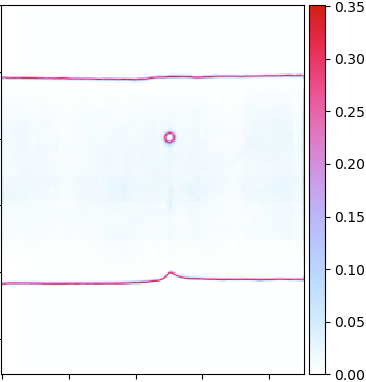}
            \caption[]%
            {{\small BCNN uncertainty.}}
        \end{subfigure}
        \hskip0.3em
        \begin{subfigure}[t]{0.3\textwidth}
            \centering 
            \includegraphics[width=0.9\textwidth]{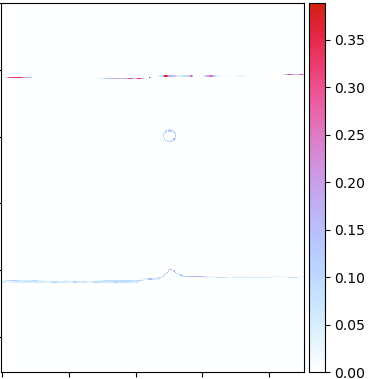}
            \caption[]%
            {{\small MCDN uncertainty.}}
        \end{subfigure}
        \caption{Results on Laser Weld Test Set Sample S4, Slice 372. While our BCNN segmentation underestimates the size of the void, it expresses a thick uncertainty band reflecting its correct size. Also, our BCNN uncertainty better captures the continuity of the edges of the weld.}
        \label{fig:S4}
\end{figure}

Figure \ref{fig:S33} shows a successful segmentation and uncertainty measurements on the S33 sample from the Laser Weld dataset. In particular, our BCNN captures the geometric variation of uncertainty around the central void. Figure \ref{fig:S4} shows another successful segmentation and uncertainty measurements on the S4 sample from the Laser Weld dataset. Note that our BCNN uncertainty map expresses consistent uncertainty measurements across the borders of the weld.

Table \ref{tab:stats} lists a selection of descriptive statistics regarding model performance on both CT scan datasets. Our BCNN achieves a higher segmentation accuracy than the MCDN on the numbered Graphite samples, but slightly lower accuracy on the named Graphite samples and the Laser Weld dataset. The labels of the CT scans were hand-segmented and are known to contain inaccuracies, especially at particle boundaries. As such, we conclude that the accuracy performance of our BCNN is similar to that of the MCDN with respect to these labels, with further assessments against refined labels left for future work.

\subsection{Validation}

Validation of UQ results is a difficult subject with no standard practice for determining whether a model's UQ is justified given the dataset. In validating our BCNN, the most relevant work in this area is due to \cite{Mukhoti}. They define two desiderata for quality uncertainty maps: a high probability of being accurate when the model is certain, denoted \( P(A|C) \), and a high probability of being uncertain when the model is inaccurate, denoted \( P(U|I) \). They estimate these quantities by evaluating accuracy and uncertainty by sliding a square patch across the image; if the patch accuracy or uncertainty is above a certain threshold, the entire patch is labeled accurate or uncertain, respectively. Our BCNN consistently outperforms the MCDN in both conditional probabilities.

In addition, \cite{Mukhoti} defines a metric called Patch Accuracy vs. Patch Uncertainty (PAvPU), which includes the above two desiderata but implicitly penalizes patches which are simultaneously accurate and uncertain.  If \(n\) represents the total number of patches, \(n_{ac}\) represents the number of patches which are accurate and certain, and \(n_{iu}\) represents the number of patches which are inaccurate and uncertain, the PAvPU score is equal to \( \frac{1}{n} (n_{ac} + n_{iu}) \) \cite{Mukhoti}.

\begin{table}[ht!]
\begin{subtable}{0.35\textwidth}
\begin{tabular}{lccc}
\toprule
Sample & Method & Accuracy & UQ Mean\tablefootnote{$\times 10^{-2}$} \\
\midrule
I & MCDN & $0.8290$ & $0.7463$ \\
  & BCNN & $0.8458$ & $7.999$ \\
\midrule
III & MCDN & $0.7413$ & $0.6665$ \\
    & BCNN & $0.7568$ & $7.618$ \\
\midrule
IV & MCDN & $0.6911$ & $0.7648$ \\
   & BCNN & $0.7220$ & $7.884$ \\
\midrule
GCA2K & MCDN & $0.9736$ & $0.2516$ \\
        & BCNN & $0.9732$ & $3.925$ \\
\midrule
25R6 & MCDN & $0.9377$ & $0.2753$ \\
     & BCNN & $0.9253$ & $4.634$ \\
\midrule
E35 & MCDN & $0.8930$ & $0.3073$ \\
    & BCNN & $0.8703$ & $4.832$ \\
\midrule
Litarion & MCDN & $0.9257$ & $0.2750$ \\
         & BCNN & $0.9140$ & $4.302$ \\
\bottomrule
\end{tabular}
\subcaption{Graphite Test Set}
\end{subtable}%
\hskip7.5em
\begin{subtable}{0.35\textwidth}
\begin{tabular}{lccc}
\toprule
Sample & Method & Accuracy & UQ Mean\tablefootnote{$\times 10^{-3}$} \\
\midrule 
S1 & MCDN & $0.9957$ & $0.7732$ \\
   & BCNN & $0.9947$ & $6.582$ \\
\midrule
S4 & MCDN & $0.9944$ & $0.9074$ \\
   & BCNN & $0.9928$ & $8.412$ \\
\midrule
S15 & MCDN & $0.9986$ & $1.032$ \\
    & BCNN & $0.9925$ & $13.09$ \\
\midrule
S26 & MCDN & $0.9863$ & $0.8058$ \\
    & BCNN & $0.9938$ & $8.865$ \\
\midrule
S31 & MCDN & $0.9977$ & $0.9090$ \\
    & BCNN & $0.9898$ & $6.629$ \\
\midrule
S32 & MCDN & $0.9921$ & $1.512$ \\
    & BCNN & $0.9890$ & $8.613$ \\
\midrule
S33 & MCDN & $0.9940$ & $1.540$ \\
    & BCNN & $0.9885$ & $7.240$ \\
\bottomrule
\end{tabular}
\subcaption{Laser Weld Test Set}
\end{subtable}
\vskip0.5\baselineskip
\caption{Test Set Statistics. Our BCNN has roughly the same accuracy performance as the MCDN, a nontrivial accomplishment because previous implementations sacrifice some accuracy for better UQ \cite{Shridhar,Ovadia}. Our BCNN also measures an order of magnitude more uncertainty than the MCDN, suggesting that our BCNN captures certain uncertainties which the MCDN cannot.}
\label{tab:stats}
\end{table}

\begin{table}[b!]
\hskip-0.75em
\begin{subtable}{0.35\textwidth}
\begin{tabular}{lcccc}
\toprule
Sample & Method & \( P(A|C) \) & \( P(U|I) \) & PAvPU3D \\
\midrule
I & MCDN & $0.9224$ & $0.4849$ & $\bm{0.7335}$ \\
 & BCNN & $\bm{0.9264}$ & $\bm{0.5485}$ & $0.6273$ \\
\midrule
III & MCDN & $0.8335$ & $0.3720$ & $\bm{0.6978}$ \\
 & BCNN & $\bm{0.8573}$ & $\bm{0.5221}$ & $0.6347$ \\
\midrule
IV & MCDN & $0.7780$ & $0.3403$ & $\bm{0.6580}$ \\
 & BCNN & $\bm{0.8017}$ & $\bm{0.4518}$ & $0.5947$ \\
\midrule
GCA2K & MCDN & $0.9997$ & $0.8830$ & $\bm{0.8764}$ \\
 & BCNN & $\bm{0.9998}$ & $\bm{0.9621}$ & $0.7752$ \\
\midrule
25R6 & MCDN & $\bm{0.9846}$ & $0.5257$ & $\bm{0.8655}$ \\
 & BCNN & $0.9807$ & $\bm{0.6445}$ & $0.7575$ \\
\midrule
E35 & MCDN & $\bm{0.9554}$ & $0.4722$ & $\bm{0.8457}$ \\
 & BCNN & $0.9320$ & $\bm{0.4971}$ & $0.7332$ \\
\midrule
Litarion & MCDN & $\bm{0.9793}$ & $0.5186$ & $\bm{0.8659}$ \\
 & BCNN & $0.9735$ & $\bm{0.6081}$ & $0.7672$ \\
\bottomrule
\end{tabular}
\subcaption{Graphite Test Set}
\end{subtable}%
\hskip7em
\begin{subtable}{0.35\textwidth}
\begin{tabular}{lcccc}
\toprule
Sample & Method & \( P(A|C) \) & \( P(U|I) \) & PAvPU3D \\
\midrule
S1 & MCDN & $1.0$ & $0.9555$ & $\bm{0.9773}$ \\
 & BCNN & $1.0$ & $\bm{1.0}$ & $0.7864$ \\
\midrule
S4 & MCDN & $1.0$ & $0.9381$ & $\bm{0.9727}$ \\
 & BCNN & $1.0$ & $\bm{1.0}$ & $0.7638$ \\
\midrule
S15 & MCDN & $1.0$ & $\bm{0.8755}$ & $\bm{0.9547}$ \\
 & BCNN & $1.0$ & $0.8300$ & $0.8093$ \\
\midrule
S26 & MCDN & $1.0$ & $0.9243$ & $\bm{0.9700}$ \\
 & BCNN & $1.0$ & $\bm{1.0}$ & $0.8541$ \\
\midrule
S31 & MCDN & $\bm{1.0}$ & $1.0$ & $\bm{1.0}$ \\
 & BCNN & $0.9998$ & $1.0$ & $0.9998$ \\
\midrule
S32 & MCDN & $\bm{1.0}$ & $1.0$ & $\bm{1.0}$ \\
 & BCNN & $0.9998$ & $1.0$ & $0.9998$ \\
\midrule
S33 & MCDN & $1.0$ & $0.9991$ & $\bm{0.9692}$ \\
 & BCNN & $1.0$ & $\bm{1.0}$ & $0.8652$ \\
\bottomrule
\end{tabular}
\centering
\subcaption{Laser Weld Test Set}
\end{subtable}
\vskip0.5\baselineskip
\caption{PAvPU3D Validation Results. Our BCNN consistently outperforms the MCDN in the conditional probabilities, implying that our BCNN better encodes the relationship between uncertainty and accuracy. Our BCNN underperforms in the PAvPU3D metric as it is penalized for being simultaneously accurate and uncertain; however, this is misleading because they are not mutually exclusive.}
\label{tab:pavpu}
\end{table}

In our CT scan datasets, the boundary between classes is often unclear due to artifacts and noise present in the scans. As such, the human labels used to quantify accuracy are inconsistent and cannot be used to properly calibrate our uncertainty. We are including the results of the PAvPU metric because, to our knowledge, it is the only relevant metric in the literature. However, it is insufficient for the validation of our UQ in the absence of ground truth labels. We leave the development of a more relevant uncertainty metric as future work.

In order to apply the PAvPU metric to our work, we introduce PAvPU3D, an extension of PAvPU calculated similarly but with a cubic patch sliding across the entire volume. We implement PAvPU3D to assess our uncertainty results using a \( 2\times 2 \times 2\) patch with accuracy threshold \( 7/8 \) and uncertainty threshold equal to the mean of the uncertainty map. We detail our results in Table \ref{tab:pavpu}.  Note that PAvPU3D includes the same implicit penalty for patches which are uncertain but accurate with respect to untrusted labels, which may not necessarily be a detrimental characteristic of the UQ. This penalty is the largest contributor to the results where the MCDN achieves a better result than our BCNN: our BCNN simply measures \textit{more} uncertainty than the MCDN. Both PAvPU and PAvPU3D reward a network which is not simultaneously uncertain and accurate; however, in the Bayesian view, uncertainty and accuracy are not mutually exclusive because uncertainty quantifies the proximity of a sample to the training distribution rather than confidence in a correct segmentation.

\subsection{Advantages for Material Simulations} \label{material-advantages}

\begin{figure}[t!]
        \centering
        \begin{subfigure}[t]{0.225\textwidth}
            \centering 
            \includegraphics[width=0.875\textwidth]{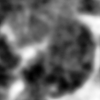}
            \caption[]%
            {{\small CT scan slice.}}
        \end{subfigure}
        \hskip0.2em
        \begin{subfigure}[t]{0.225\textwidth}  
	  \centering 
            \includegraphics[width=\textwidth]{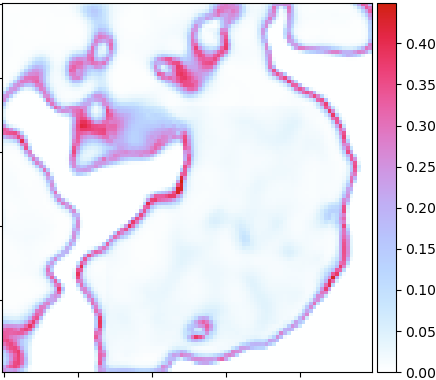}
            \caption[]%
            {{\small BCNN 33-67 uncertainty.}}
        \end{subfigure}
        \hskip0.2em
        \begin{subfigure}[t]{0.225\textwidth}    
		\centering 
            \includegraphics[width=\textwidth]{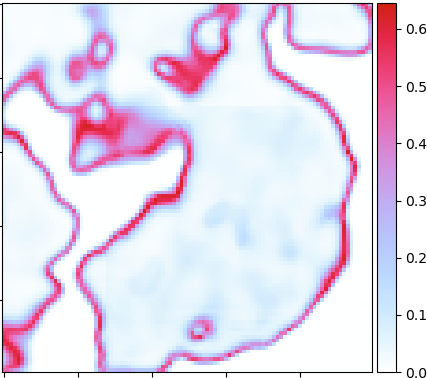}
            \caption[]%
            {{\small BCNN 20-80 uncertainty.}}   
        \end{subfigure}
		\begin{subfigure}[t]{0.225\textwidth}    
		\centering 
            \includegraphics[width=\textwidth]{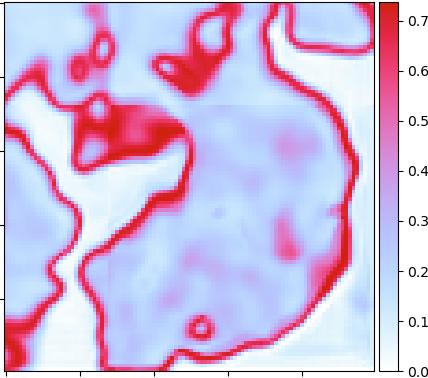}
            \caption[]%
            {{\small BCNN 5-95 uncertainty.}}   
        \end{subfigure}
        \vskip0.33\baselineskip
        \begin{subfigure}[t]{0.225\textwidth}
		\centering
            \includegraphics[width=0.875\textwidth]{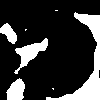}
            \caption[]%
            {{\small Target segmentation.}}
        \end{subfigure}
        \hskip0.2em
        \begin{subfigure}[t]{0.225\textwidth}   
            \centering 
            \includegraphics[width=\textwidth]{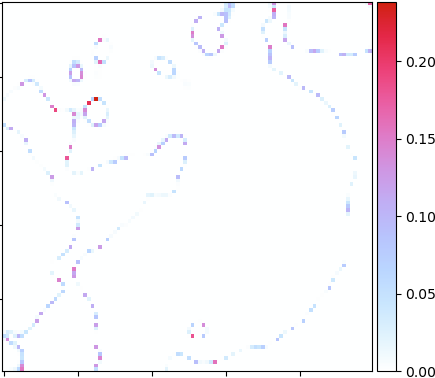}
            \caption[]%
            {{\small MCDN 33-67 uncertainty.}}    
        \end{subfigure}
        \hskip0.2em
        \begin{subfigure}[t]{0.225\textwidth}  
            \centering 
            \includegraphics[width=\textwidth]{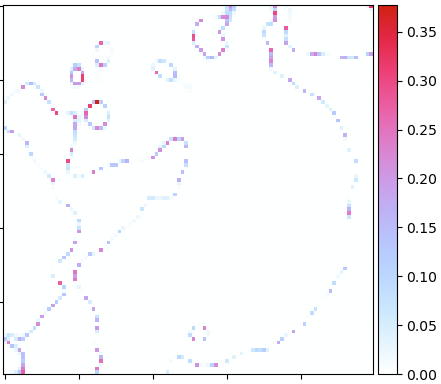}
            \caption[]%
            {{\small MCDN 20-80 uncertainty.}}
        \end{subfigure}
	  \begin{subfigure}[t]{0.225\textwidth}    
		\centering 
            \includegraphics[width=\textwidth]{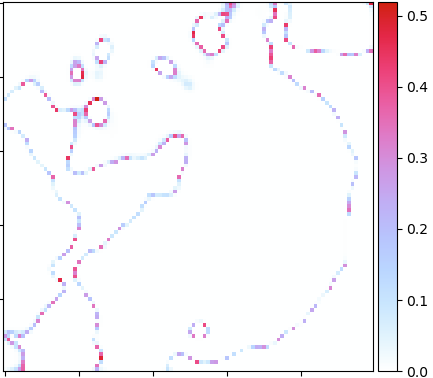}
            \caption[]%
            {{\small MCDN 5-95 uncertainty.}}   
        \end{subfigure}
        \caption{Zoomed Uncertainty Maps on Graphite Test Set Sample 25R6, Slice 272. Each uncertainty map is calculated as the difference between the specified sigmoid percentiles. Our BCNN introduces  additional areas of uncertainty as the interval grows larger, suggesting that it captures a meaningful uncertainty distribution, while the MCDN uncertainty maps barely change.}
        \label{fig:confidence}
\end{figure}

The objective of performing UQ on materials datasets is to obtain uncertainties which can inform and propagate throughout simulations involving said materials. For example, when simulating the performance of a sample from the Graphite dataset to bound its various physical properties, it is crucial to know the contact points of the material. The uncertainty maps generated by our BCNN are by construction the difference of an interval of sigmoid percentiles -- these represent confidence intervals on the segmentation, so we can infer the probability of a certain contact point occurring in the CT scanned material. These probabilities can be integrated into simulations to obtain confidence intervals on the physical properties of the materials.

The voxel-wise nature of the MCDN uncertainty maps produce very jagged, unrealistic confidence intervals with little physical meaning. In contrast, the continuity and visual gradients of the uncertainty map generated by our BCNN enable better approximations to the actual geometric uncertainty in both the Graphite and Laser Weld materials. Our BCNN allows us to smoothly probe the uncertainty when performing simulations and justify each error bound we obtain with interpretable uncertainty maps, a major advantage when performing simulations for high-consequence scenarios. In Figure \ref{fig:confidence} we provide an example of probing the uncertainty of our model at different confidence intervals; our BCNN captures geometric differences between intervals which the MCDN cannot. This suggests that our BCNN learns a distribution of uncertainty which is closely linked to the geometry of the individual material, while the MCDN uncertainty is a voxel-wise measurement unrelated to the local properties of the CT scan. In practice, this means that our BCNN can better detect uncertainties in the geometry of the segmentation and therefore provide more credible confidence intervals during simulation.

\section{Conclusion}

Our BCNN is the first deep learning method which generates statistically credible geometric uncertainty maps and the first Bayesian CNN which robustly scales with three-dimensional data. By measuring uncertainty in the weight space, our BCNN provides interpretable uncertainty quantification for segmentation problems. We present UQ results on CT scans of graphite electrodes and laser-welded metals used in material simulations where accurate geometric UQ is critical. Our BCNN achieves equal or better segmentation accuracy than MCDNs in most cases and outperforms MCDNs in recent uncertainty metrics \cite{Mukhoti}, and we show that probing our BCNN uncertainty leads to more credible confidence intervals on the segmentation than the MCDN. Future investigation will likely include extending our BCNN to semantic segmentation, isolating aleatoric and epistemic uncertainties, and integrating our results with other UQ techniques such as deep ensembles \cite{Lak}.

\section{Acknowledgments}
We'd like to thank Kellin Rumsey for his advice on measuring the uncertainty of our models, Carter Jameson for his suggestion to use group normalization, and Dustin Tran for his suggestion to use KL annealing. We would also like to thank Kyle Karlson for providing the Laser Weld dataset and Chance Norris for curating and segmenting the Graphite dataset.

\clearpage
% ---- Bibliography ----
%
% BibTeX users should specify bibliography style 'splncs04'.
% References will then be sorted and formatted in the correct style.
%
\bibliographystyle{splncs04}
\bibliography{bcnn}

\clearpage
\begin{appendices}
\renewcommand{\thesection}{\appendixname~\Alph{section}}

\section{Chunking and Prediction}

Given a specified chunk size such as \( (88 \times 172 \times 172) \), our algorithm decomposes an extremely large CT scan volume (for example, of dimension \( (742 \times 1100 \times 1100) \)) into chunks of that size. The user can also specify the step -- that is, how much overlap the chunks should have with each other on the original volume. A step size of 1 is no overlap, 2 is an overlap of \( 1/2 \), 3 an overlap of \( 2/3 \), and so on. We recommend a step size of 3. The output of our chunking algorithm is a 5-dimensional volume with chunks arranged consecutively in a list, a python dictionary of the top left coordinate of each chunk in the original volume, and the shape of the original volume. Algorithm \ref{alg:chunk} provides pseudocode for this algorithm.

Once we have the decomposed volume, we need to perform predictions on each chunk with the neural network and stitch the output back together. Since our BCNN has a stochastic forward pass, we allow the user to specify how many samples they would like to run. This value must be divisible by the batch size. The algorithm sums the predictions and stores a counts volume of how many times each individual voxel was predicted, then divides the sums by the counts. It also keeps track of the sigmoid volume (\textit{i.e.}, the mean predictions) and different percentiles (\textit{e.g.}, the 33\textsuperscript{rd} and 67\textsuperscript{th} percentile predictions). Finally, we calculate a prediction volume, which entails casting each voxel in the sigmoid volume to either 0 (if the sigmoid value is below $0.5$) or 1 (if the sigmoid value is above $0.5$), and an uncertainty volume obtained by subtracting the 33\textsuperscript{rd} percentile from the 67\textsuperscript{th}. Notice that these percentile volumes are usable as confidence intervals on the prediction as described in Section \ref{material-advantages}. Algorithm \ref{alg:pred} provides pseudocode for this algorithm.

However, Algorithm \ref{alg:pred} can cause chunking artifacts in the prediction volume. We hypothesize that the artifacts occur at chunk boundaries due to the model lacking context of the area. Thus, we set a threshold on the boundary of the chunks and discard those predictions. This technique is easible integrable with Algorithm \ref{alg:pred}, since the counts volume easily accounts for the voxels which were dropped in a certain chunk. The implementation of this subroutine, shown in Algorithm \ref{alg:rmv} using a threshold of 10\%, mainly consists of case-checking to ensure that we do not divide by zero in the counts volume.

See our codebase at \url{https://github.com/sandialabs/bcnn} for python implementations of each of these algorithms.

\begin{algorithm}[H]
\caption{Chunking Algorithm}
\label{alg:chunk}
\begin{algorithmic}[1]
\Require orig-volume, chunk-size, step
\State new $\gets$ [], coords $\gets$ [], shape $\gets$ orig-volume.shape
\State $z_{max} \gets$ shape[0] $-$ chunk-size[0]
\State $x_{max} \gets$ shape[1] $-$ chunk-size[1]
\State $y_{max} \gets$ shape[2] $-$ chunk-size[2]
\If {$z_{max} < 0\text{ or }x_{max} < 0\text{ or }y_{max} < 0$}
\State Return "Volume too small for given chunk size"
\EndIf
\State $z_{flag} \gets False, x_{flag} \gets False, y_{flag} \gets False$
\For {$z \text{ in } range(0, \text{shape[0]}, \text{chunk-size[0]} // \text{step})$}
\State $x_{flag} \gets False, y_{flag} \gets False$
\If {$z_{flag}$}
\State Break
\ElsIf {$z > z_{max}$}
\If {$z_{max} = 0 \text{ or } z_{max} \text{ mod } (\text{chunk-size[0]} // \text{step}) = 0$}
\State Break
\EndIf
\State $z \gets z_{max}$
\State $z_{flag} \gets True$
\EndIf
\For {$x \text{ in } range(0, \text{shape[1]}, \text{chunk-size[1]} // \text{step})$}
\State $y_{flag} \gets False$
\If {$x_{flag}$}
\State Break
\ElsIf {$x > x_{max}$}
\If {$x_{max} = 0 \text{ or } x_{max} \text{ mod } (\text{chunk-size[1]} // \text{step}) = 0$}
\State Break
\EndIf
\State $x \gets x_{max}$
\State $x_{flag} \gets True$
\EndIf
\For {$y \text{ in } range(0, \text{shape[2]}, \text{chunk-size[2]} // \text{step})$}
\If {$y_{flag}$}
\State Break
\ElsIf {$y > y_{max}$}
\If {$y_{max} = 0 \text{ or } y_{max} \text{ mod } (\text{chunk-size[2]} // \text{step}) = 0$}
\State Break
\EndIf
\State $y \gets y_{max}$
\State $y_{flag} \gets True$
\EndIf
\State $\text{coords.append}((z, x, y))$
\State $\text{new.append}(\text{orig-volume}[z:z+\text{chunk-size[0]}, x:x+\text{chunk-size[1]}, y:y+\text{chunk-size[2]}, :])$
\EndFor
\EndFor
\EndFor
\State Return new, coords, shape
\end{algorithmic}
\end{algorithm}

\begin{algorithm}[H]
\caption{Prediction Algorithm}
\label{alg:pred}
\begin{algorithmic}[1]
\Require model, volume, test-shape, chunk-shape, coords, num-samples, batch-size
\State $\text{sigmoid} \gets zeros(\text{test-shape})$
\State $\text{counts} \gets zeros(\text{test-shape})$
\State $\text{percentiles} \gets \text{[}zeros(\text{test-shape}) * 2\text{]}$
\State $\text{percentile-points} \gets \text{[}33, 67\text{]}$
\For {$i, (\text{chunk, coord}) \text{ in } enumerate(zip(\text{volume, coords}))$}
\State $\text{chunk-samples} \gets zeros((\text{num-samples},) + \text{chunk-shape})$
\State Add an empty batch dimension to chunk in the first axis
\State $\text{batch} \gets repeat(\text{chunk}, \text{batch-size}, axis=0)$
\For {$j\text{ in }range(0, \text{num-samples, batch-size})$}
\State $\text{chunk-samples}\text{[}j:j+\text{batch-size]} \gets model.predictbatch(\text{batch})$
\EndFor
\State Increment counts by one at each voxel contained in the chunk position in the original volume, calculated using chunk-shape and coord.
\State $\text{chunk-mean} \gets mean(\text{chunk-samples}, axis=0)$
\State Increment counts by the corresponding voxel in chunk-mean at each voxel contained in the chunk position in the original volume, calculated using chunk-shape and coord.
\State $\text{percentile-samples} \gets percentile(\text{chunk-samples, percentile-points}, axis=0)$
\State Increment percentiles by the corresponding voxel in percentile-samples at each voxel contained in the chunk position in the original volume, calculated using chunk-shape and coord. Do this for each percentile in percentile-points.
\EndFor
\State $\text{sigmoid} \gets \text{sigmoid} / \text{counts}$
\State $\text{percentiles} \gets \text{percentiles} / \text{counts}$
\State $\text{pred} \gets \text{sigmoid}.copy()$
\State $\text{pred}[\text{pred} > 0.5] \gets 1.$
\State $\text{pred}[\text{pred} \leq 0.5] \gets 0.$
\State $\text{unc} \gets \text{percentiles[1]} - \text{percentiles[0]}$
\State Return pred, unc, percentiles
\end{algorithmic}
\end{algorithm}

\begin{algorithm}[H]
\caption{Subroutine for Artifact Removal}
\label{alg:rmv}
\begin{algorithmic}[1]
\Require This subroutine belongs between lines 11-12 in Algorithm \ref{alg:pred}. Require the previously instantiated variables test-shape, chunk-shape, coord, and chunk-samples.
\State $\text{trimmed-shape} \gets \text{chunk-shape}$
\For {$i \text{ in } [0,1,2]$}
\State $\text{border} \gets ceil(\text{chunk-shape[$i$]} * 0.1)$
\If {$\text{coord[$i$]} \neq 0 \text{ and coord[$i$]} \neq \text{test-shape[$i$]} - \text{chunk-shape[$i$]}$}
\State $\text{chunk-samples[$i+1$]} \gets \text{chunk-samples[$i+1$][border:-border]}$
\State $\text{coord[$i$]} \gets \text{coord[$i$] $+$ border}$
\State $\text{trimmed-shape[$i$]} \gets \text{trimmed-shape[$i$]} - (2 * \text{border})$
\ElsIf {$\text{coord[$i$]} \neq 0$}
\State $\text{chunk-samples[$i+1$]} \gets \text{chunk-samples[$i+1$][border:]}$
\State $\text{coord[$i$]} \gets \text{coord[$i$] $+$ border}$
\State $\text{trimmed-shape[$i$]} \gets \text{trimmed-shape[$i$]} - \text{border}$
\ElsIf {$\text{coord[$i$]} \neq \text{test-shape[$i$]} - \text{chunk-shape[$i$]}$}
\State $\text{chunk-samples[$i+1$]} \gets \text{chunk-samples[$i+1$][:-border]}$
\State $\text{trimmed-shape[$i$]} \gets \text{trimmed-shape[$i$]} - \text{border}$
\EndIf
\EndFor
\State Use trimmed-shape instead of chunk-shape in Line 12, 14, and 16 in Algorithm \ref{alg:pred}
\end{algorithmic}
\end{algorithm}

\end{appendices}

\end{document}